\documentclass[aps,prl,twocolumn,groupedaddress,,showpacs]{revtex4}
\usepackage{graphicx}
\usepackage{dcolumn}
\usepackage{bm}
\usepackage{amsmath}
\usepackage{amssymb}
\usepackage{color}

\definecolor{yellow}{rgb}{1,1,0}
\definecolor{pink}{rgb}{0.68,0.92,1}
\definecolor{lightgreen}{rgb}{1,0.6,0.78}

\begin{document}

\title{Experimental implementation of adiabatic passage between different topological orders}

\author{Xinhua Peng$^{1,2}$}
\email{xhpeng@ustc.edu.cn}
\author{Zhihuang Luo$^{1}$, Wenqiang Zheng$^{1}$}
\author{Supeng Kou$^3$}
\email{spkou@bnu.edu.cn}
\author{Dieter Suter$^4$}
\author{Jiangfeng Du$^{1,2}$}
\email{djf@ustc.edu.cn}
\affiliation{$^1$Hefei National Laboratory for Physical Sciences at Microscale and Department of Modern Physics, University of Science and Technology of China, Hefei, Anhui 230026, China}
\affiliation{$^{2}$Synergetic Innovation Center of Quantum Information $\&$ Quantum Physics,
University of Science and Technology of China, Hefei, Anhui 230026, China}
\affiliation{$^3$Department of Physics, Beijing Normal University, Beijing, 100875, China}
\affiliation{$^4$Fakult\"{a}t Physik, Technische Universit\"{a}t Dortmund, 44221, Dortmund, Germany}

\begin{abstract}
Topological orders are exotic phases of matter existing in strongly correlated quantum systems, which are beyond the usual symmetry description and
cannot be distinguished by local order parameters. Here we report an experimental quantum simulation of the Wen-plaquette spin model with different topological orders in a nuclear magnetic resonance system, and observe the adiabatic transition between two $Z_2$ topological orders through a spin-polarized phase by measuring the nonlocal closed-string (Wilson loop) operator. Moreover, we also measure the entanglement properties of the topological orders.
This work confirms the adiabatic method for preparing topologically ordered states and provides an experimental tool for further studies of complex quantum systems.
\end{abstract}

\pacs{03.65.Ud, 64.70.Tg, 03.67.Lx, 76.60.-k}

\maketitle

Over the past 30 years, it has become increasingly clear that the Landau symmetry-breaking theory cannot describe all phases of matter
and their quantum phase transitions (QPTs) \cite{QPT,Landau1937,Ginzburg1950}.
The discovery of the fractional quantum Hall (FQH) effect \cite{TSG8259} indicates the existence of an exotic state of matter termed topological orders \cite{Wen1990TO}, which are beyond the usual symmetry description.
This type of orders has some interesting properties, such as robust ground state degeneracy that depends on the
surface topology \cite{Wen1990TO2}, quasiparticle fractional statistics \cite{Arovas}, protected edge states \cite{Wen1995}, topological entanglement entropy \cite{k3} and so on. Besides the importance in condensed matter physics, topological orders have also been found
potential applications in fault-tolerant topological quantum computation \cite{k1,Nayak2008,Stern2013}. Instead of naturally occurring physical systems (e.g., FQH), two-dimensional spin-lattice models, including the toric-code model \cite{k1},
the Wen-plaquette model \cite{Wen2003}, and the Kitaev model on a hexagonal lattice \cite{k2}, were found to exhibit $Z_2$ topological orders.
The study of such systems therefore  provides an opportunity to understand more features of topological orders and the associated topological QPTs \cite{p3,HammaPRL2008,HammaPRB2008}. A large body of theoretical work exists on these systems, including several proposals for their physical implementation in cold atoms \cite{Duan2003}, polar molecules \cite{Micheli2006} or arrays of Josephson junctions \cite{You2010}.
However, only a very small number of experimental investigations have actually demonstrated such topological properties
(e.g., anyonic statistics and robustness) using either photons \cite{Pan2012} or nuclear spins \cite{Du2007}. However, in these experiments, specific entangled states having topological properties have been dynamically generated, instead of direct Hamiltonian engineering and ground-state cooling which are extremely demanding experimentally.

Rather than the toric-code model, the first spin-lattice model with topological orders, here we study an alternative exactly solvable spin-lattice model -- the Wen-plaquette model \cite{Wen2003}.
Two different $Z_2$ topological orders exist in this system; their stability depends on the sign of the coupling constants of the four-body interaction.
Between these two phases, a new kind of phase transition occurs when the couplings vanish.
So far, neither these topological orders nor this topological QPT have been observed experimentally.
The two major challenges are (i) to engineer and to experimentally control  complex quantum systems with four-body interactions and
(ii) to detect efficiently the resulting topologically ordered phases.
Along the lines suggested by Feynman \cite{Feynman1982}, complex quantum systems can be efficiently simulated on quantum simulators,
i.e., programmable quantum systems whose dynamics can be efficiently controlled.
Some earlier experiments have been studied, e.g., in condensed-matter physics \cite{Pan2012,Du2007,Peng2005}
and quantum chemistry \cite{Lu2011} (see the review on quantum simulation \cite{QSreview} and references therein).
Quantum simulations thus offer the possibility to investigate strongly correlated systems exhibiting topological orders and other complex quantum systems that are challenging for simulations on classical computers.

In this Letter, we demonstrate an experimental quantum simulation of the Wen-plaquette model in a nuclear magnetic resonance (NMR) system and observe an adiabatic transition between two different topological orders that are separated by a spin-polarized state.
To the best of our knowledge,  this is the first experimental observation of such a system based on using the Wilson loop operator,
which corresponds to a nonlocal order parameter of a topological QPT \cite{HammaPRL2008,HammaPRB2008}.
Both topological orders are further confirmed to be highly entangled by quantum state tomography.
The experimental adiabatic method paves the way towards constructing and initializing a topological quantum memory \cite{Dennis2002,Jiang2008}.

\begin{figure}[tbp]
\centering \includegraphics[width=8cm]{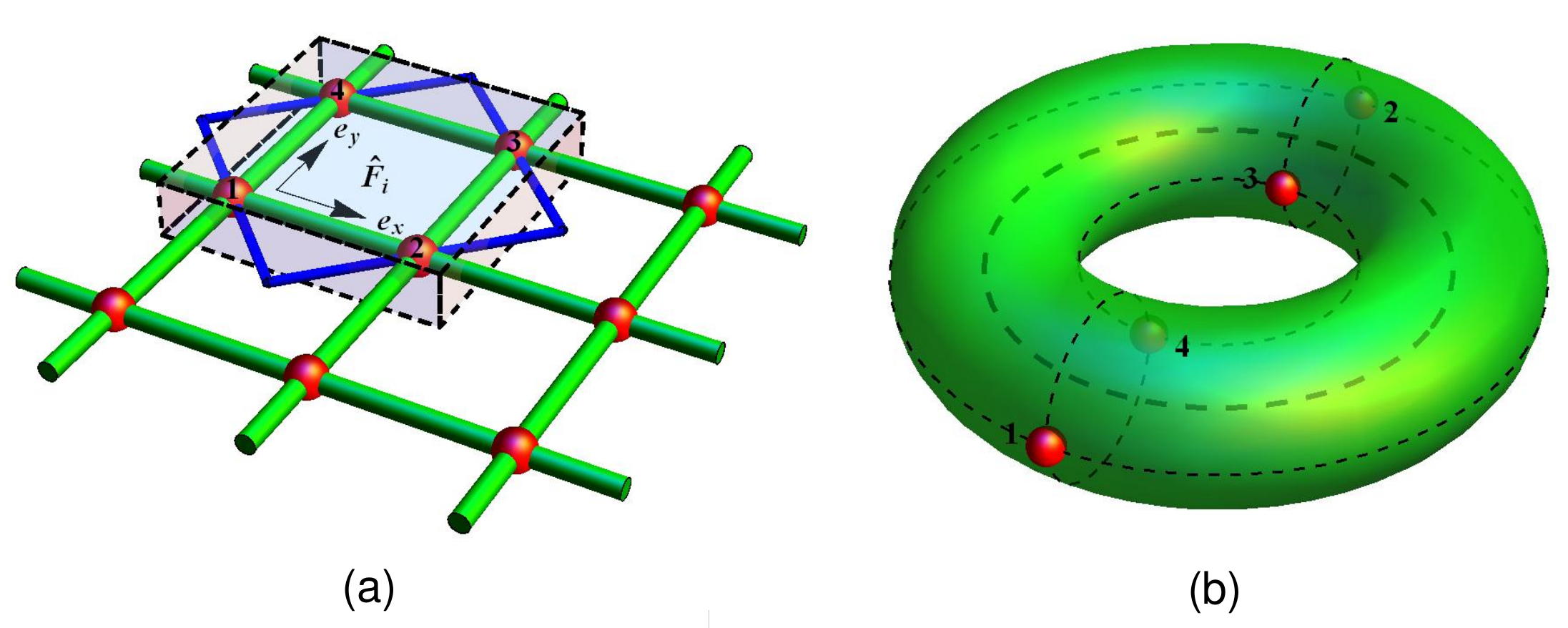}
\caption{(Color online.) (a) Wen-plaquette model. The red spheres represent
spin-$1/2$ particles. The plaquette operator is $\hat{F_i}=\hat{\protect%
\sigma}_{i}^{x}\hat{\protect \sigma}_{i+\hat{e}_{x}}^{y}\hat{\protect \sigma%
}_{i+\hat{e}_{x}+\hat{e}_{y}}^{x}\hat{\protect \sigma}_{i+\hat{e}%
_{y}}^{y}$. The closed string (blue) represents the Wilson loop in a $2 \times 2$ lattice.
(b) A torus formed from a $2 \times 2$ lattice (labeled by 1,2 3,4) with  periodic
boundary condition. }
\label{fig:lattices}
\end{figure}

We focus on the Wen-plaquette model  \cite{Wen2003} shown in Fig.~\ref{fig:lattices}(a),
an exactly solvable quantum spin model with  $Z_{2}$ topological orders. It is described by the Hamiltonian
\begin{equation}
\hat{H}_{Wen}=-J\sum_{i}\hat{F_{i}},
\end{equation}%
where $\hat{F_{i}}=\hat{\sigma}_{i}^{x}\hat{\sigma}_{i+\hat{e}_{x}}^{y}%
\hat{\sigma}_{i+\hat{e}_{x}+\hat{e}_{y}}^{x}\hat{\sigma}_{i+\hat{e}%
_{y}}^{y}$ is the plaquette operator that acts on the four spins surrounding
a plaquette.
Since $\hat{F_{i}}^{2}=1$, the eigenvalues of $\hat{F_{i}}$ are
$F_{i}=\pm 1$. We see that when $J>0$ the ground state has all $F_{i}=1$ and
when $J<0$ the ground state has all $F_{i}=-1$.  According to
the classification of the projective symmetry group \cite{Wen2003}, they correspond to two
types of topological orders: $Z_{2}A$ and $Z_{2}B$ order, respectively. It is obvious that
both topological orders have the same global symmetry as that belongs to the
Hamiltonian. So one cannot use the concept of \textquotedblleft
spontaneous symmetry breaking" and the local order parameters to distinguish
them. In $Z_{2}A$\emph{\ }$(Z_{2}B)$ order, a ``magnetic vortex" (or $m$-particle)
is defined as $F_{i}=-1$ ($F_{i}=1$) at an even sub-plaquette and an
``electric charge" (or $e$-particle) is $F_{i}=-1$ ($F_{i}=1$) at an odd sub-plaquette \cite{WenPRD2003}. Due to
the mutual semion statistics between $e$- and $m$-particles, their bound states obeys
fermionic statistics  \cite{k2,WenPRD2003}.
Physically, in $Z_{2}B$ topological order, a fermionic
excitation (the bound state of $e$ and $m$) sees a $\pi $-flux tube around each
plaquette and acquires an Aharonov--Bohm phase $e^{i\pi }$
when moving around a plaquette, while in $Z_{2}A$ topological order, the
fermionic excitation feels no additional phase when moving around each
plaquette. Thus the transition at $J=0$ represents a new kind of phase transition
that changes quantum orders but not symmetry \cite{Wen2003,WenPRD2003}.

\begin{figure}[tbp]
\centering \includegraphics[width=8cm]{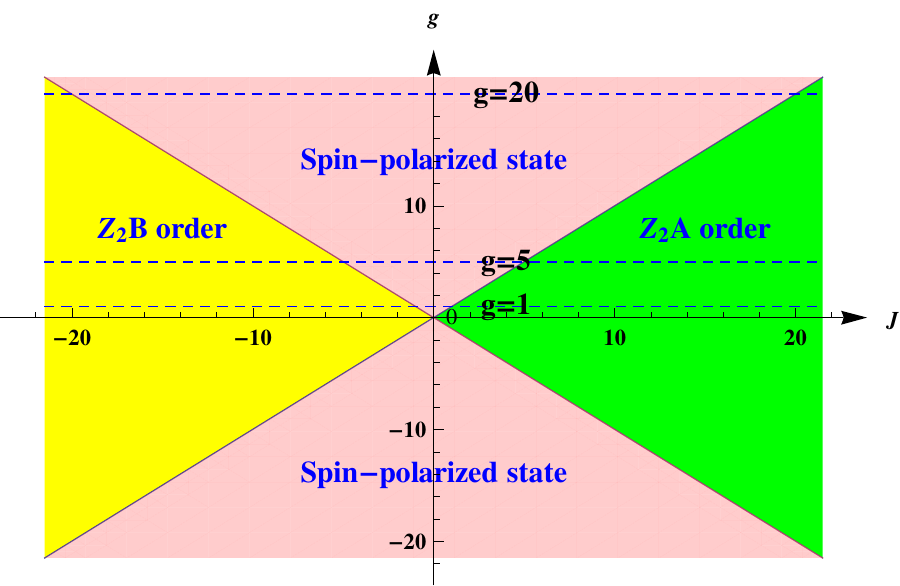}
\caption{(Color online.) 2D phase diagram of Wen-plaquette model in a
transverse field. The yellow, green and pink regions represent $Z_2B$, $Z_2A$
topological orders, and the spin-polarized state, respectively. The three
dashed lines correspond to the tested values of $g=1,g=5$ and $g=20$ in the
experiment.}
\label{fig:phase diagram}
\end{figure}

However, it is difficult to directly observe the transition from $Z_{2}A$ to $Z_{2}B$ topological order in the experiment,
because the energies of all quantum states are zero at the critical point.
Instead, the Wen-plaquette model in a transverse field
\begin{equation}
\hat{H}_{tol}=\hat{H}_{Wen}-g\sum_{i}\hat{\sigma _{i}^{x}}.
\label{Htol}
\end{equation}%
is often studied \cite{p3,HammaPRL2008,HammaPRB2008}. Without loss of generality, we consider the case $g>0$.
Figure \ref{fig:phase diagram} shows its 2D phase diagram, which contains
three regions in which the ground state is $Z_{2}B$ order when $J\ll -g$, $Z_{2}A$ order
when $J\gg g$ and a spin-polarized state without topological order when $%
|J|\ll g$, respectively.
From Fig. \ref{fig:phase diagram}, we can see that
by changing $J$, the ground state of the system is driven from $Z_{2}A$ to $%
Z_{2}B$ topological order through the trivial spin-polarized state. The spin-polarized region  from one topological order to the other one depends
on the size of the transverse field strength $g$: the smaller $g$ is, the
narrower the region of spin-polarized state becomes.
If $g$ vanishes (or $J$ is large enough), a QPT occurs between the two types of topological orders  \cite{Wen2003}.
The above results are valid only for infinite systems.
For finite systems, the situation is  more complicated. For example,
the topological degeneracies of the system depend on the type of the lattice (even-by-even, even-by-odd, odd-by-odd lattices).
However, the properties of the topological orders persist in the Wen-plaquette model with finite-size lattices \cite{Wen2003}.

The simplest finite system that exhibits topological orders consists of a $2\times 2$
lattice with periodic boundary condition, as shown in Fig.~\ref{fig:lattices}%
(b). The Hamiltonian can be described as
\begin{equation}
\hat{H}_{Wen}^{4}=-2J(\hat{\sigma}_{1}^{x}\hat{\sigma}_{2}^{y}\hat{\sigma}%
_{3}^{x}\hat{\sigma}_{4}^{y}+\hat{\sigma}_{1}^{y}\hat{\sigma}_{2}^{x}\hat{%
\sigma}_{3}^{y}\hat{\sigma}_{4}^{x}).
\end{equation}%
The fourfold degeneracy of the ground
states is a topological degeneracy and the two ground
states for $J<0$ and for $J > 0$ have different quantum
orders \cite{Wen2003}.
Adding a transverse field, we obtain the transverse Wen-plaquette model $\hat{H}_{tol}$ in Eq.  \eqref{Htol} for the finite system,
where the degeneracy is partly lifted \cite{p3}.
For the case $g>0$, the non-degenerate ground state is:
\begin{equation}
|\psi _{g}\rangle \approx \left \{
\begin{array}{ll}
|\psi _{Z_{2}B}\rangle =|\phi ^{+}\rangle _{13}|\phi ^{+}\rangle _{24}, &
J\ll -g<0 \\
|\psi _{SP}\rangle =|++++\rangle , & J=0 \\
|\psi _{Z_{2}A}\rangle =|\psi ^{+}\rangle _{13}|\psi ^{+}\rangle _{24}, &
J\gg g>0%
\end{array}%
\right. .  \label{gs_H2}
\end{equation}%
Here $|\phi ^{+}\rangle =\frac{1}{\sqrt{2}}(|00\rangle +|11\rangle )$, $%
|\psi ^{+}\rangle =\frac{1}{\sqrt{2}}(|01\rangle +|10\rangle )$ and $%
|+\rangle =\frac{1}{\sqrt{2}}(|0\rangle +|1\rangle )$.
The energy-level diagram and the ground state are given in Ref. \cite{SM}.
Eq. (\ref{gs_H2}) shows that both topological orders are symmetric and possess bipartite entanglement,
while the spin-polarized state $|\psi _{SP}\rangle $ is a product
state without entanglement.

\begin{figure}[tbp]
\centering \includegraphics[width=8cm]{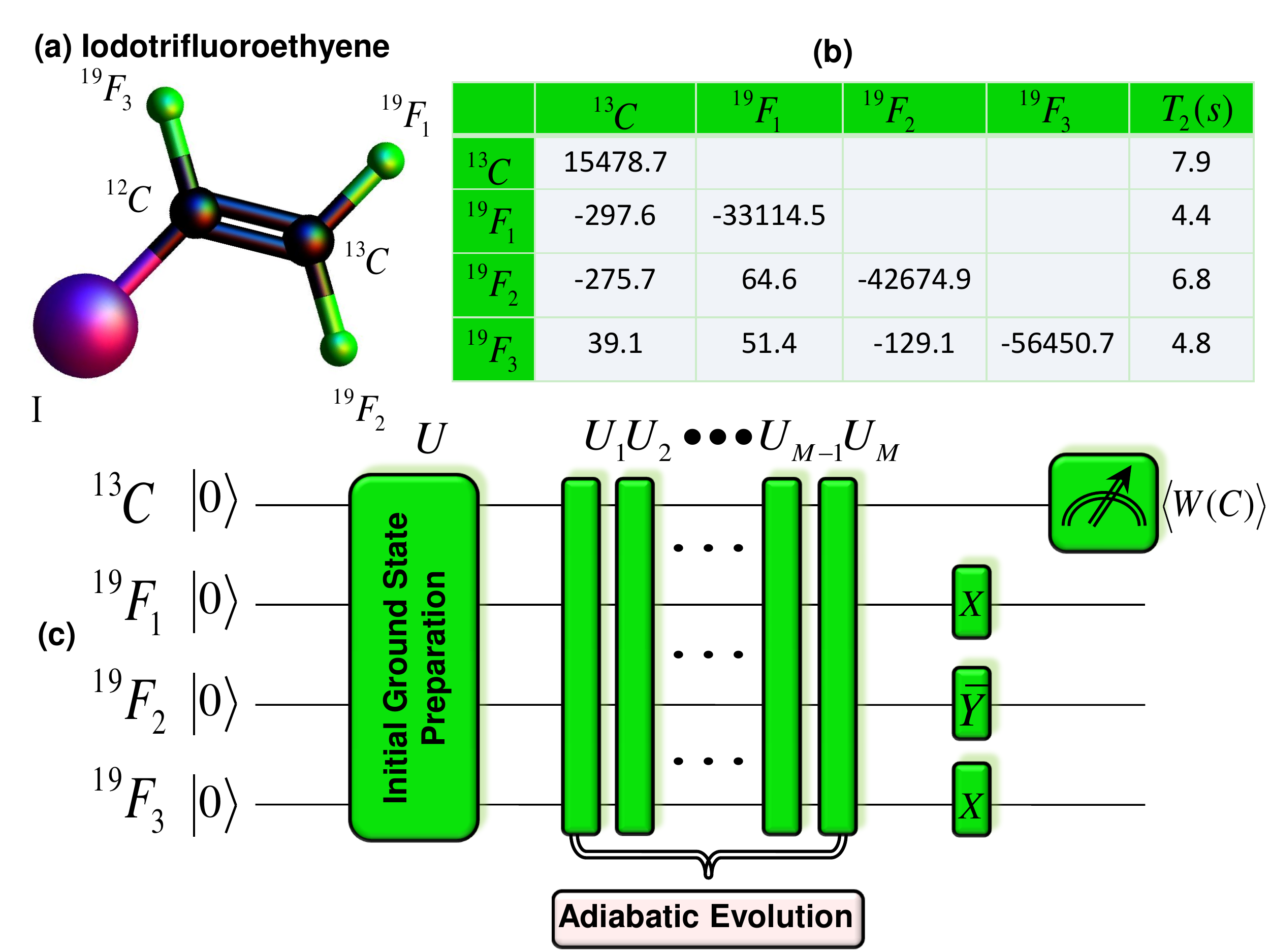}
\caption{(Color online.) (a) Molecular structure of lodotrifluroehtylene.
One $^{13}$C and three $^{19}$F nuclei are used as a four-qubit quantum
simulator. (b) Relevant parameters measured at $T=300 K$. The diagonal and
nondiagonal elements represent the chemical shifts and the coupling
constants in units of Hz, respectively. The measured spin-lattice relaxation
times ($T_1$) are  $21$ s for $^{13}$C and $12.5$ s for $^{19}$F.
(c) Quantum circuit for observing the topological-order-transition in the Wen-plaquette model.
$X$ and $\bar{Y}$ represent $\frac{\protect \pi}{2}$ rotations of single
qubits around the $x$ and $-y$ axes, respectively. }
\label{fig:sample and quantum circuit}
\end{figure}

The  physical four-qubit system we used in the experiments consists of the nuclear spins
in Iodotrifluroethylene (C$_2$F$_3$I) molecules with one $^{13}$C  and three $^{19}$F nuclei. 
Figure ~\ref{fig:sample and quantum circuit} (a) and (b) show its molecular structure and relevant properties \cite{SM}.
The natural Hamiltonian of this system in the doubly rotating frame is
\begin{equation}
\hat{H}_{NMR}=\sum_{i=1}^{4}\frac{\omega _{i}}{2}\hat{\sigma}
^{z}_{i}+\sum_{i<j,=1}^{4}\frac{\pi J_{ij}}{2}\hat{\sigma} ^{z}_{i}\hat{%
\sigma} ^{z}_{j},
\end{equation}%
where $\omega _{i}$ represents the chemical shift of spin $i$ and $J_{ij}$ the coupling
constant. The experiments were carried out on a Bruker
AV-400 spectrometer ($9.4T$) at room temperature $T=300$ K. The
 temperature fluctuation was controlled to $<0.1$ K, which results in a frequency stability within $1$ Hz.
 Figure \ref{fig:sample and quantum circuit}(c) shows the quantum circuit for the experiment, which can be divided into three steps:
($i$) preparation of the initial ground state of the Hamiltonian $\hat{H}_{tol}[J(0)]$ for a given transverse field $g$,
($ii$)  adiabatic simulation of $\hat{H}_{tol}[J(t)]$ by changing the
control parameter $J$ from $J(0)$ to $J(T)$, and
($iii$)  detection of the resulting state.

To prepare the system in the ground state, we used the technique of
pseudo-pure states (PPS): $\hat{\rho}_{\psi}=\frac{1-\epsilon}{16}\mathbf{I}%
+\epsilon| \psi \rangle \langle \psi |$, with $\mathbf{I}$ representing the $%
16\times16$ identity operator and $\epsilon \approx10^{-5}$ the
polarization. Starting from the thermal state, we prepared the PPS $\hat{\rho%
}_{0000}$ by line-selective pulses \cite{Peng2001}. 
The experimental fidelity of $\hat{\rho}_{0000}$ defined by $|Tr(\hat{\rho}%
_{th}\hat{\rho}_{exp})|/\sqrt{Tr(\hat{\rho}_{th}^2)Tr(\hat{\rho}_{exp}^2)}$
was around $97.7\%$. Then we obtained the initial ground state $\hat{\rho}%
_{\psi_g}$ of $\hat{H}_{tol}[J(0)] $ by a unitary operator realized by a
GRAPE pulse  \cite{Glaser2005} with a duration of 6 ms.

To observe the ground-state transition, we implemented an adiabatic transfer from $\hat{H}_{tol}[J(0)]$
to $\hat{H}_{tol}[J(T)]$ \cite{Messiah1976}.
The sweep control parameter $J(t)$ was numerically optimized and implemented
as a discretised scan with $M$ steps :
\begin{equation}
\hat{U}_{ad}=\prod_{m=1}^{M}\hat{U}_{m}[J_{m}]=\prod_{m=1}^{M}e^{-i\hat{H}%
_{tol}[J_{m}]\tau },
\end{equation}%
where the duration of each step is $\tau =T/M$. The adiabatic limit corresponds to
$T,M\rightarrow \infty ,\tau \rightarrow 0$.
Using $M=31$, the optimized  sweep  reaches a theoretical fidelity $> 99.5$\%
of the final state with respect to the true ground state.
For each step of the adiabatic passage, we designed the NMR pulse sequence to create an effective Hamiltonian, i.e., $\hat{H}_{tol}[J_m]$ \cite{SM}.

\begin{figure}[tbp]
\centering \includegraphics[width=8cm]{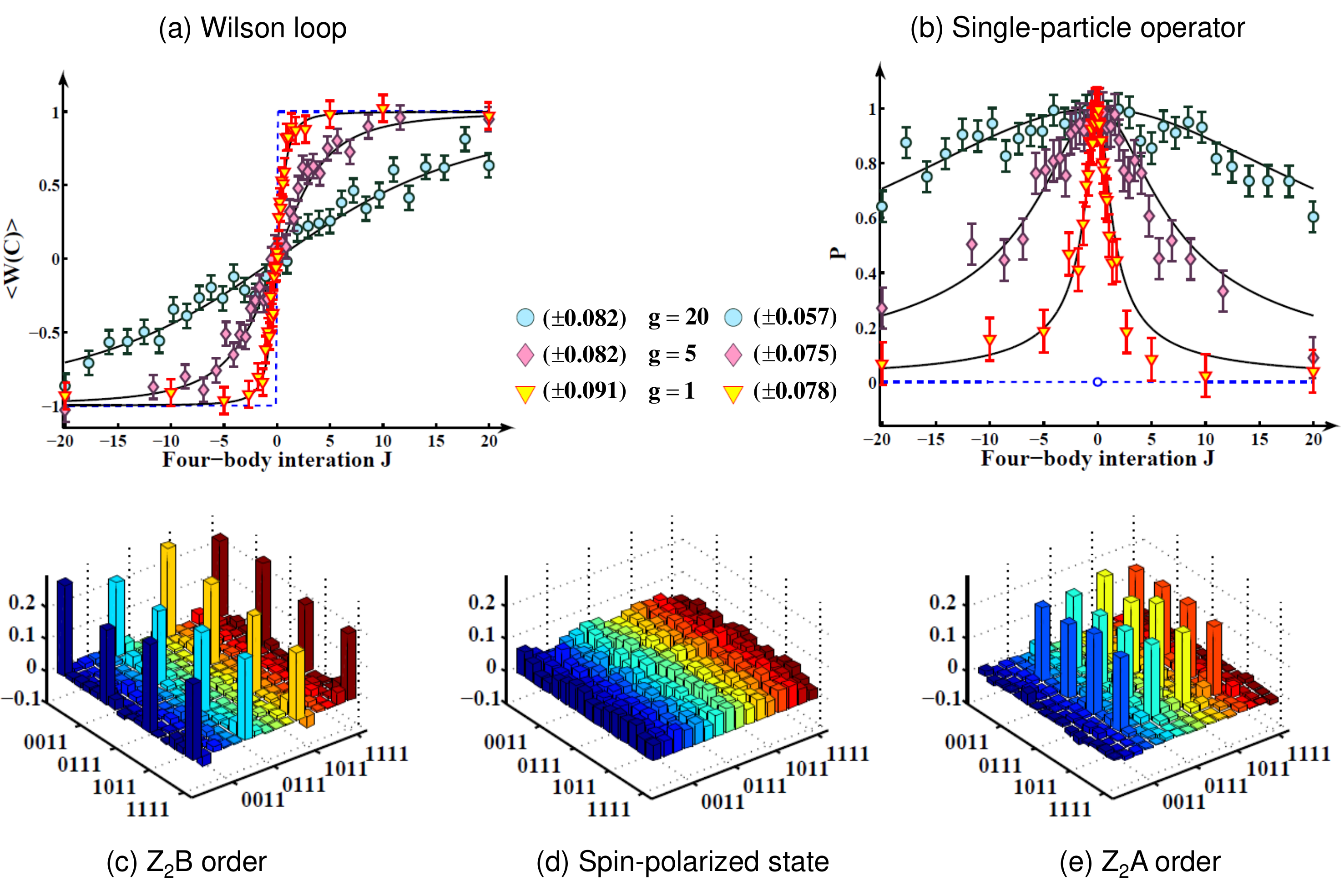}
\caption{(Color online.) (a) Measured expectation values of the nonlocal
string operator -- the Wilson loop $\langle \hat{W}(C) \rangle$. (b)
Measured values $P$ of the local single-particle operator. The experimental
points are denoted by the symbols {$\triangledown$}, {$\lozenge$} and {$%
\circ $} for $g=1,g=5$ and $g=20$, respectively, along with the theoretical
expectations denoted by the black solid lines.
The error bars indicate the standard deviations of the experimental measurements.
The sharp transition denoted by the blue dashed line is the
theoretical expectation when $g= 0 $. (c) (d) and (e) Real parts of
experimentally reconstructed density matrices for the ground states with $g
= 1$ at $J = -20$, $J = 0$ and $J = 20$, corresponding to $Z_2B$
topologically ordered state $\vert \protect \psi_{Z_2B} \rangle$,
spin-polarized state $\vert \protect \psi_{SP}\rangle$ and $Z_2A$
topologically ordered state $\vert \protect \psi_{Z_2A} \rangle$. All
imaginary parts of the density matrices are small. The rows and columns
represent the standard computational basis in binary order, from $\vert 0000
\rangle$ to $\vert 1111 \rangle$.}
\label{fig:results}
\end{figure}

In the experiment, we employed the Wilson loop \cite{HammaPRL2008,HammaPRB2008,Kogut1979} to detect the transition
between two different topological orders. The effective theory of topological orders is a $Z_2$ gauge theory and the observables must be gauge invariant quantities. The Wilson loop operator is gauge invariant and can be as a nonlocal order parameter. It is defined as $\hat{W}(C)=\prod_{C}\hat{%
\sigma} _{i}^{\alpha _{i}} $, where the product $\prod_{C}$ is over all
sites on the closed string $C$, $\alpha _{i}=y$ if $i$ is even and $\alpha
_{i}=x$ if $i$ is odd \cite{WenPRD2003}.
For the $2 \times 2$ lattice system, this corresponds to
$\hat{W}(C)=\hat{\sigma}_{1}^{x}\hat{\sigma}_{2}^{y}\hat{\sigma}_{3}^{x}\hat{\sigma}_{4}^{y}$.
The experimental results of $\langle \hat{W}%
(C)\rangle$ can be obtained by recording the carbon spectra after a
read-out pulse $[\frac{\pi }{2}]_{x}^{F_{1}}[\frac{\pi }{2}]_{\bar{y}%
}^{F_{2}}[\frac{\pi }{2}]_{x}^{F_{3}}$.
Figure \ref{fig:results}(a) shows the resulting data for three sets of  experiments with
$g=1,$ $g=5$, $g=20$ and $J$ varying from $-20 $ to $20$.
When $|J/g| \gg 1$, $\langle \hat{W}%
(C)\rangle $ is close to $\pm 1$, corresponding to $Z_{2}A$/$Z_{2}B$
topological order.
The results shown in figure \ref{fig:results}(a) verify that
the transition region becomes narrower and sharper as $g$ decreases.
In the absence of the transverse field, $g\to0$, the ground state makes a sudden transition  at $J=0$
from  $Z_{2}B$ to  $Z_{2}A$ topological order.
This is a novel QPT between different
topological orders  \cite{Wen2003}. These results also show that the Wilson loop is a
useful nonlocal order parameter that  characterises the different $Z_2$
topological orders very well.

To demonstrate more clearly  that this topological QPT goes beyond
Landau symmetry-breaking theory and cannot be described by local order
parameters, we also measured the single-particle operator of the $^{13}$C
spin:
\begin{equation*}
P = | Tr[\hat{\rho}_{f} ( \hat{\sigma}_1^x - i \hat{\sigma}_1^y)]| = \sqrt{
Tr(\hat{\rho}_f \hat{\sigma}_1^x)^2 + Tr(\hat{\rho}_f \hat{\sigma}_1^y)^2 }.
\end{equation*}
This was performed by measuring the magnitude of the $^{13}$C NMR signal
while decoupling $^{19} $F. Here $\hat{\rho}_f$ is the final  state
at the end of the adiabatic scan. Due to the symmetry of the Hamiltonian, the
values of $P$ are equal for all the four spins.
Figure \ref{fig:results}(b) shows the experimental results.
They are symmetric with
respect to $J=0$, which means that the different $Z_2$ topological orders
cannot be distinguished by the local order parameter.

By performing complete quantum state tomography \cite{Lee}, we reconstructed
the density matrices for $Z_{2}B$ order ($J=-20$), for  the spin-polarized state ($J=0
$) and  $Z_{2}A$ order ($J=20$) for  $g=1$. The real parts of
these density matrices are shown in Fig.\ref{fig:results}(c), (d) and (e).
The experimental fidelities are  $95.2\%,$ $95.6\%$ and $95.7\%$,
respectively. From these reconstructed density matrices, we also calculated the entanglement: for both topological orders, $C(\hat{\rho}_{13}^{exp})\approx C(\hat{\rho}_{24}^{exp})\approx 0.89$, while the others were close to 0; for the spin-polarized state, all $C(\hat{\rho}_{ij}^{exp})$ are almost zero. Here $\hat{\rho}_{ij}^{exp}$ is the reduced density matrix of two spins $i, j$ obtained by partially tracing out the other spins from the experimentally reconstructed density matrix $\hat{\rho}^{exp}$
and the concurrence is defined as $C(\hat{\rho}_{ij}^{exp})=max\{\lambda_1-\lambda_2-\lambda_3-\lambda_4, 0\}$, where $\lambda_k$s (in decreasing order) are the square roots of the eigenvalues of $\hat{\rho}_{ij}^{exp}(\hat{\sigma}^y_i\hat{\sigma}^y_j)\hat{\rho}_{ij}^{exp^{\ast}}(\hat{\sigma}^y_i\hat{\sigma}^y_j)$\cite{Wootters1998}. Therefore, the topological orders exhibit the same bipartite entanglement between qubits 1, 3 and 2, 4 in agreement with Eq. (\ref{gs_H2}). These experimental
results are in good agreement with theoretical expectations.
The relatively minor deviations can be attributed mostly to  the imperfections of
the GRAPE pulses, the initial ground state preparation and the spectral
integrals \cite{SM}.

Instead of studying naturally existing topological phases like those in
quantum Hall systems, lattice-spin models can be designed to exhibit
interesting topological phases.
One example is the Wen-plaquette model, which includes many-body interactions.
Such interactions have not been found in naturally occurring systems, but
they can be generated as effective interactions in quantum simulators.
Using an NMR quantum simulator, we provide a first proof-of-principle experiment that
implements an adiabatic transition between two
different $Z_{2}$ topological orders through a spin-polarized state in the
transverse Wen-plaquette model.
Such models are beyond Landau symmetry-breaking
theory and cannot be described by local order parameters. Ref. \cite{HammaPRB2008}
presented a numerical study of a QPT from a spin-polarized to a
topologically ordered phase using a variety of previously proposed QPT
detectors and demonstrated their feasibility.
Furthermore, we also demonstrated in an experiment  that the nonlocal Wilson loop operator
can be a nontrivial detector of topological QPT between different topological orders.
This phenomenon requires further investigation to be properly understood.

Although a $2\times2$ lattice is a very small finite-size system, topological orders exist in the Wen-plaquette model with periodic lattice of finite size \cite{Wen2003}.
The validity of the quantum simulation of the topological orders in such a small system also comes from the fairly short-range spin-spin correlations. When $|g/J| \le 1$,  all quasi-particles (the electric charges,
magnetic vortices and fermions) perfectly localize that leads to zero spin-spin correlation length \cite{k1,Yu2009}.
Therefore the topological properties of the ground state persist in such a small system, including the topological degeneracy, the statistics of the quasi-particles and the non-zero Wilson loop \cite{SM}.
The present method can in principle be expanded to larger systems with more spins,
which allows one to explore more interesting physical phenomena,
such as lattice-dependent topological degeneracy \cite{Wen2003}, quasiparticle fractional statistics \cite{Arovas,k2}
and the robustness of the ground state degeneracy against local perturbations \cite{Wen1990TO2,HammaPRL2008,HammaPRB2008,Yu2009}.
Quantum simulators using larger spin systems can be more powerful than classical computers and permit the research of topological orders and their physics
beyond the capabilities of classical computers.
Nevertheless, our present experimental results demonstrate the feasibility of small
quantum simulators for strongly correlated quantum systems, and the usefulness of
the adiabatic method for constructing and initializing a topological quantum memory.

We thanks L. Jiang and C. K. Duan for the helpful discussion. This work is supported by NKBRP (973 programs 2013CB921800, 2014CB848700, 2012CB921704, 2011CB921803), NNSFC (11375167, 11227901, 891021005),  CAS (SPRB(B) XDB01030400), RFDPHEC (20113402110044).

\section{Supplementary Materials}
\section{Theoretical calculations in the transverse Wen-Plaquette model}
\subsection{Energy levels and ground state}

\begin{figure}[h]
\includegraphics[width=9cm]{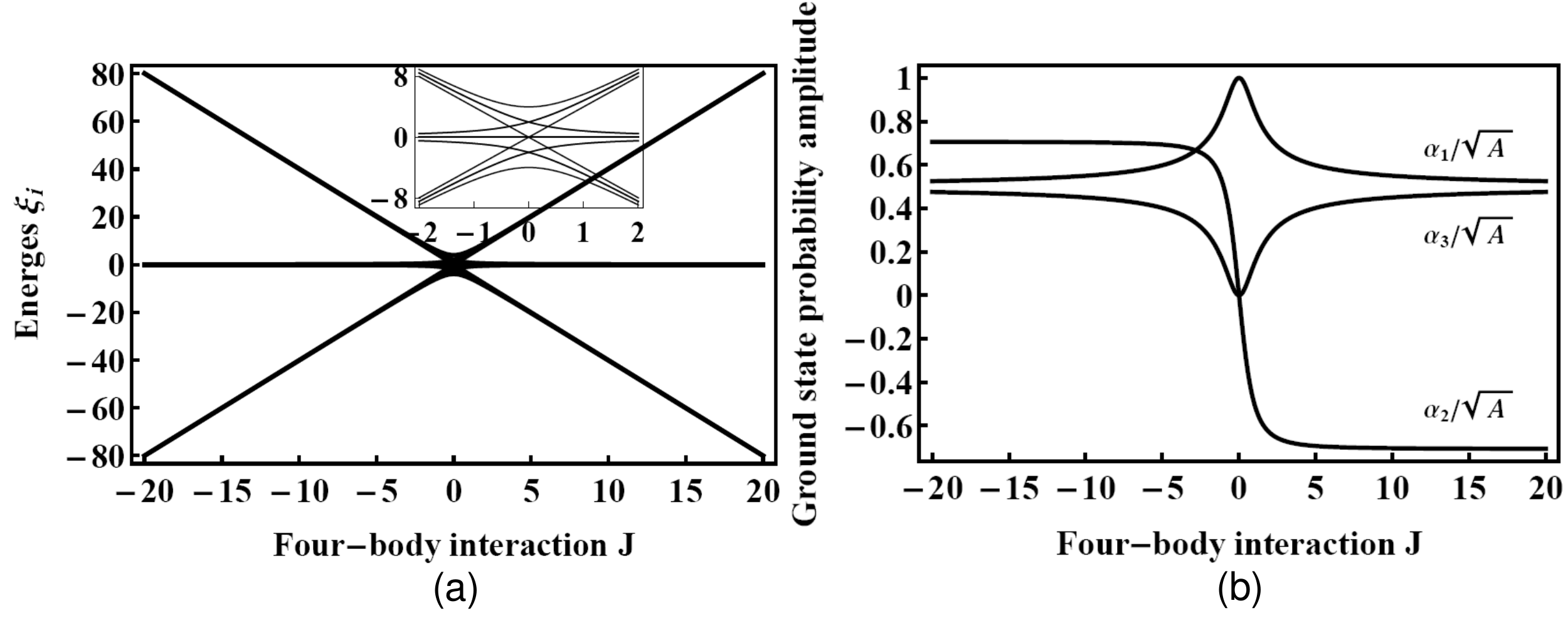}
\caption{(a) Energy-level diagram of 2 by 2 lattices in the transverse
Wen-plaquette model when $g=1$. (b) Probability amplitudes $(\protect \alpha_i/%
\protect \sqrt{A},i=1,2,3.)$ of ground state $|\protect \psi_g\rangle$ for $g=1
$.}
\label{fig:energy and ground state}
\end{figure}

With periodic boundary condition, the total Hamiltonian of 2 by 2
lattices in the transverse Wen-Plaquette model is
\begin{equation}  \label{Htol}
\hat{H}_{tol}=-2J(\hat{\sigma}_{1}^{x}\hat{\sigma}_{2}^{y}\hat{\sigma}%
_{3}^{x}\hat{\sigma}_{4}^{y}+\hat{\sigma}_{1}^{y}\hat{\sigma}_{2}^{x}\hat{%
\sigma}_{3}^{y}\hat{\sigma}_{4}^{x})-g\sum_i^4\hat{\sigma}_i^x.
\end{equation}
In the representation of $\hat{\sigma _{x}}$ basis, its ground state is
\begin{equation}
|\psi _{g}\rangle =\frac{1}{\sqrt{A}}[\alpha _{1}|0000\rangle _{x}-\alpha
_{2}\frac{|0101\rangle _{x}+|1010\rangle _{x}}{\sqrt{2}}+\alpha
_{3}|1111\rangle _{x}],  \label{psi_g} \nonumber
\end{equation}%
where $A$ is the normalization constant and $\alpha _{1}=J^{2}+2g^{2}+2g%
\sqrt{g^{2}+J^{2}},$ $\alpha _{2}=\sqrt{2}J(g+\sqrt{g^{2}+J^{2}})$ and $%
\alpha _{3}=J^{2}$. The corresponding ground-state energy is
\begin{equation}
\varepsilon =-4\sqrt{g^{2}+J^{2}}.
\end{equation}
Figure \ref{fig:energy and ground state} shows its energy-level diagram and the probability amplitudes of the ground state $|\psi
_{g}\rangle $ as a function of the four-body interaction strength $J$ for a transverse field (here we take $g=1$). The energy-level diagram is
symmetry about $J=0$ because of the symmetric transverse field. When $|J/g| \gg 1$, the ground state is progressively
four-fold degeneracy (the full four-fold degeneracy of ground state when $g = 0$ is partially lifted when $g \ne 0$, see the subplot of Fig.~ \ref{fig:energy and ground state}(a)). Note that the ground-state energy seems to be smooth due to the scale-size effect and the transverse field. For the Wen-Plaquette model (i.e. g = 0), an actual level-crossing in the four-spin system creates a point of nonanalyticity of the ground state energy as a function of the control parameter $J$. As theoretically predicted by X. Wen \cite{Wen2003}, a quantum phase transition (QPT) between two different topological orders ($Z_2A$ and $Z_2B$ orders) occurs at $J = 0$. However, the transition cannot be directly observed in experiment due to the level-crossing (the adiabatic passage will fail at the transition point). Therefore, we turn to the transverse Wen-Plaquette model (i.e., $g \ne 0$), where a second-order QPT between one topological order and spin-polarized state occurs at $J/g = \pm 1$ in the thermodynamic limit \cite{Yu2008, Feng2007, HammaPRL2008, HammaPRB2008}. Accordingly, these two topological orders ($Z_2A$ and $Z_2B$ orders) are connected by a spin-polarized state, as shown in Fig. 2 in the paper. The region of spin-polarized state will become narrow as $|g/J|$ decreases. When $g /J \to 0$,  the region turns into a point, and the ground-state transition in the Wen-Plaquette model \cite{Wen2003} can be asymptotically observed in the experiment. Therefore, as long as $g$ is small enough, the main features of the ground state in Wen-plaquette model persists (except for the point of $J = 0$). As shown in Figure~\ref{fig:energy and ground state}(b), it clearly illustrates that
there are two different types of the entangled ground states for $J\gg 1$
and $J\ll-1$.

\subsection{Spin-spin correlations}

The validity of the quantum simulation of the topological orders in the
Wen-plaquette model on $2$-by-$2$ lattice comes from the fairly short range
spin-spin correlations. For the Wen-plaquette model in the exactly solvable
limit ($g/J\rightarrow 0$) all quasi-particles (the electric charges,
magnetic vortices and fermions) have flat bands. In other words, the
quasi-particles cannot move at all. Such perfect localization of
quasi-particles leads to no spin-spin correlations for two spins on
different sites, $\langle \hat{\sigma}_{i}^{x}\hat{\sigma}_{j}^{x}\rangle
=\langle \hat{\sigma}_{i}^{y}\hat{\sigma}_{j}^{y}\rangle =\langle \hat{\sigma%
}_{i}^{z}\hat{\sigma}_{j}^{z}\rangle =0$ for $i\neq j$. Under the
perturbations, the quasi-particles begin to hop. For example, the term $%
g\sum_{i}\hat{\sigma _{i}^{x}}$ drives the quasi-particles hopping along
diagonal direction \cite{k1,KouPRL2009,Yu2009}. Therefore one may manipulate the dynamics of the
quasi-particles by adding the external field and consequently control the
spin-spin correlation length $\xi $.

\begin{figure}
  \includegraphics[width=9cm]{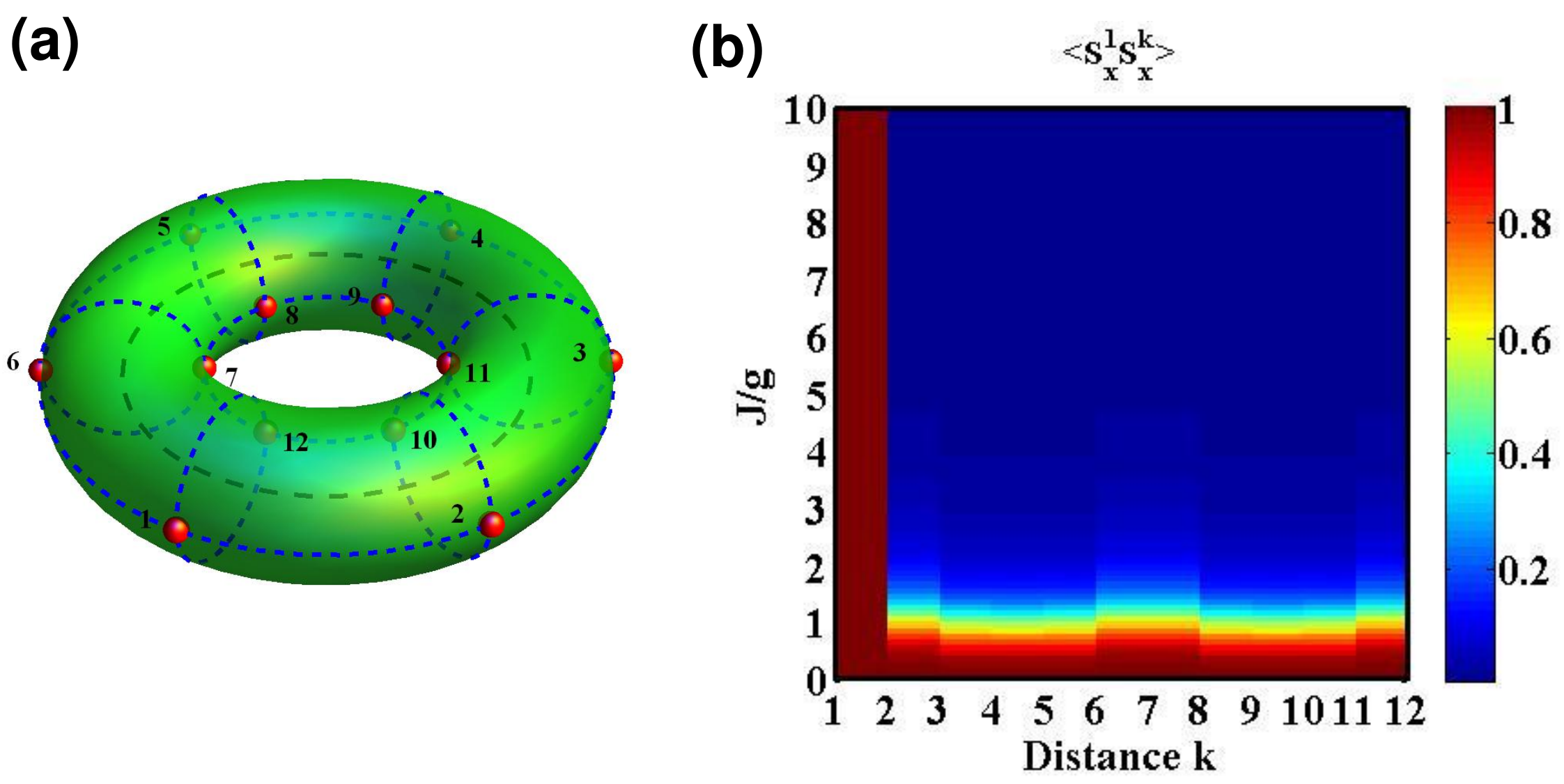}\\
  \caption{(a)The Wen-plaquette model on a $2\times 6$ lattices with periodic condition and (b)the spin-spin correlation of $\langle S_x^1S_x^k\rangle$ vs distance k and the ratio of $J/g$. }\label{fig:s2}
\end{figure}

By using the exact diagonalization technique of the Wen-plaquette model on a
$2$-by-$6$ lattices with periodic boundary condition, we obtain the
spin-spin correlations for two spins with different distances via the
strength of the external field $g$. See the results in Fig. \ref{fig:s2}. From this
figure, one can see that in the region of $g/J<1$, the spin-spin correlation
length is always smaller than $2$. As a result, for the Wen-plaquette model
on $2$-by-$2$ lattice, we can also get the topological properties including
the topological degeneracy, the statistics of the quasi-particles and the
non-zero Wilson loop. For example, the energy splitting of the degenerate
ground states is estimated by $\Delta E\sim e^{-L/\xi }$ where $L$ is size
of the system\cite{k1,KouPRL2009,Yu2009}. In the limit of $g/J\rightarrow 0$, due to
perfect localization, $\xi \rightarrow 0$, for the Wen-plaquette model on $2$%
-by-$2$ lattice the energy splitting of the degenerate ground states
disappears, $\Delta E\sim e^{-L/\xi }\rightarrow 0$ ($L=2$). However, in the
region of $g/J>1$, the ground state is spin-polarized phase without
topological order, of which the spin-spin correlation length is infinite.
Due to its trivial properties, we can also simulate the system on a lattice
of small size.

\section{Experimental procedure}

\subsection{Quantum simulator and characterization}
We chose the $^{13}$C, and three $^{19}$F nuclear spins of
iodotrifluoroethyene dissolved in d-chloroform as a four-qubit quantum
simulator. The exact characterization of the quantum simulator is very
important for precise quantum control in the experiments. The transverse
relaxation times were measured by the CPMG pulse sequence. The absolute
values of the J-coupling constants were obtained from the equilibrium
spectrum. We determined their relative sign by creating observable
three-spin orders, such as $I_{1}^{x}I_{2}^{z}I_{3}^{z}$ and measuring the
1-D NMR spectrum. This method requires a simpler pulse sequence and less
experimental time than 2D NMR sequences like $\beta $-COSY \cite{Claridge}.
Because we used an unlabeled sample, the molecules with a $^{13}$C nucleus,
which we used as the quantum register, were present at a concentration of
about $1\%$. The $^{19}$F spectra were dominated by signals from the
three-spin molecules containing the $^{12}$C isotope, while the signals from
the quantum simulator with the $^{13}$C nucleus appeared only as small ($%
0.5\%$) satellites. The accurate $^{19}$F chemical shifts are thus hidden in
the very small signals, which are obtained by exact assignments to
distinguish them from spurious molecules with a $^{13}$C nucleus.

\subsection{Adiabatic passage}

\begin{figure}[h]
\centering \includegraphics[width=8.5cm]{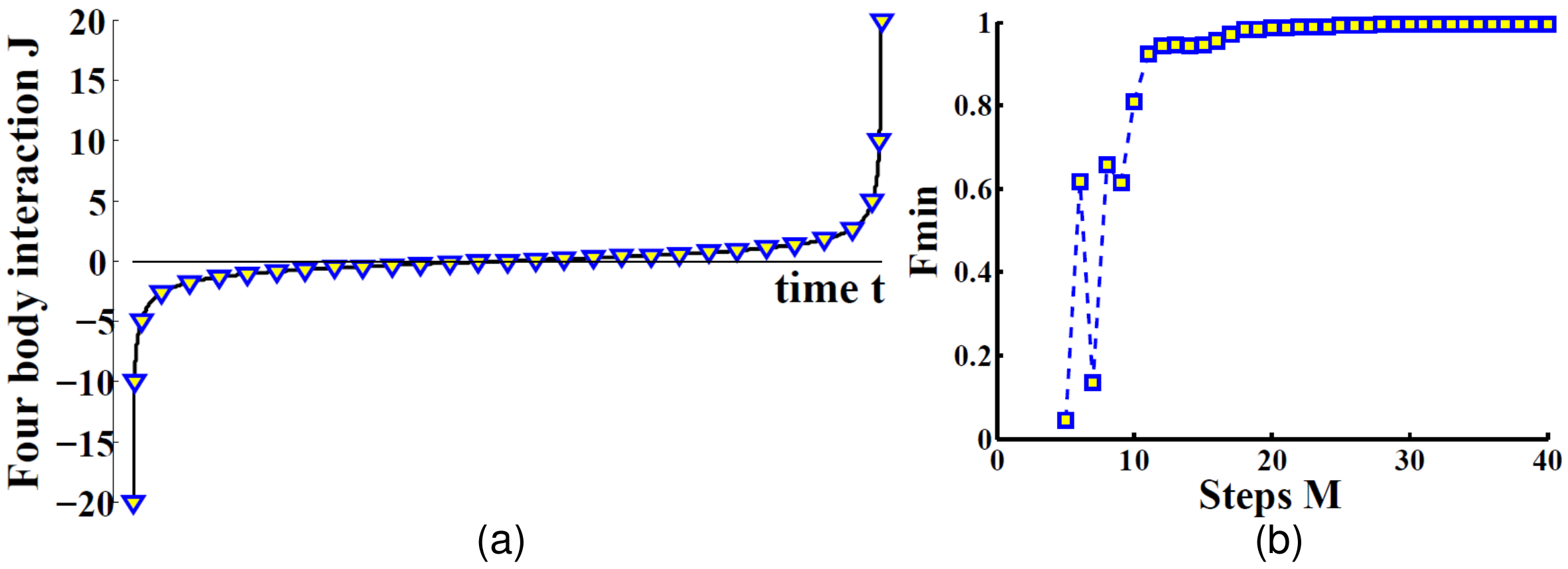}
\caption{(a) Adiabatic four-body-interaction sweep $J(t)$. The solid line
was calculated for constant adiabaticity parameter [see Eq. (5)] for a
transverse field $g=1$. The $\bigtriangledown$ points represent the $M=31$
interpolations on the solid line for the discretized scan. (b) Numerical
simulation of the minimum fidelities during the adiabatic passage vs. the
number of steps.}
\label{fig:adiabatic passage}
\end{figure}

We simulated the adiabatic transition from a topological ordered state to
another one through a spin-polarized state, where the four-body interaction $%
J$ was adiabatically driven as a control parameter. To ensure that the
system always stays in the instantaneous ground state, the variation of the
control parameter has to be sufficiently, i.e., the adiabatic condition \cite%
{Messiah1976}
\begin{equation}
\bigg|\frac{\langle \psi _{g}|\dot{\psi} _{e}\rangle }{\varepsilon
_{e}-\varepsilon _{g}}\bigg|\ll 1
\end{equation}%
is satisfied, where the index $e$ represents the excited state. The condition can be
rewritten as
\begin{equation}
\bigg|\frac{dJ(t)}{dt}\bigg|\ll 1/\bigg|\frac{\langle \psi _{g}|\frac{%
\partial \hat{H}_{tol}}{\partial J}|\psi _{e}\rangle }{(\varepsilon
_{e}-\varepsilon _{g})^{2}}\bigg|.  \label{adia}
\end{equation}%
Equation \eqref{adia} determines the optimal sweep of control parameter $J(t)
$, denoted by solid line in Fig.~\ref{fig:adiabatic passage}(a). For the experimental implementation,
we discretized the time-dependent parameter $J(t)$ into $M$ segments during the total duration of the adiabatic passage $T$.
The adiabatic condition is satisfied when both $T, M \to \infty$ and the duration of each step $\tau \to 0$.
To determine the optimal number $M$ of steps in the adiabatic transfer,
we used a numerical simulation of the minimum fidelity $F_{min}$ encountered during the scan
as a function of the number of steps into which the evolution is divided (see Fig. \ref {fig:adiabatic passage}(b)), where we fixed the total evolution time $T=6.5684$.
The fidelity is calculated as the overlap of the state with the ground state at the relevant position.
When $M=31$, the minimal fidelity is 0.995, which fully indicates the state of the system is always close to its instantaneous ground state in the whole adiabatic passage.

\subsection{Experimental Hamiltonian Simulation of the transverse Wen-plaquette model}

\begin{figure}[tbp]
\includegraphics[width=9cm]{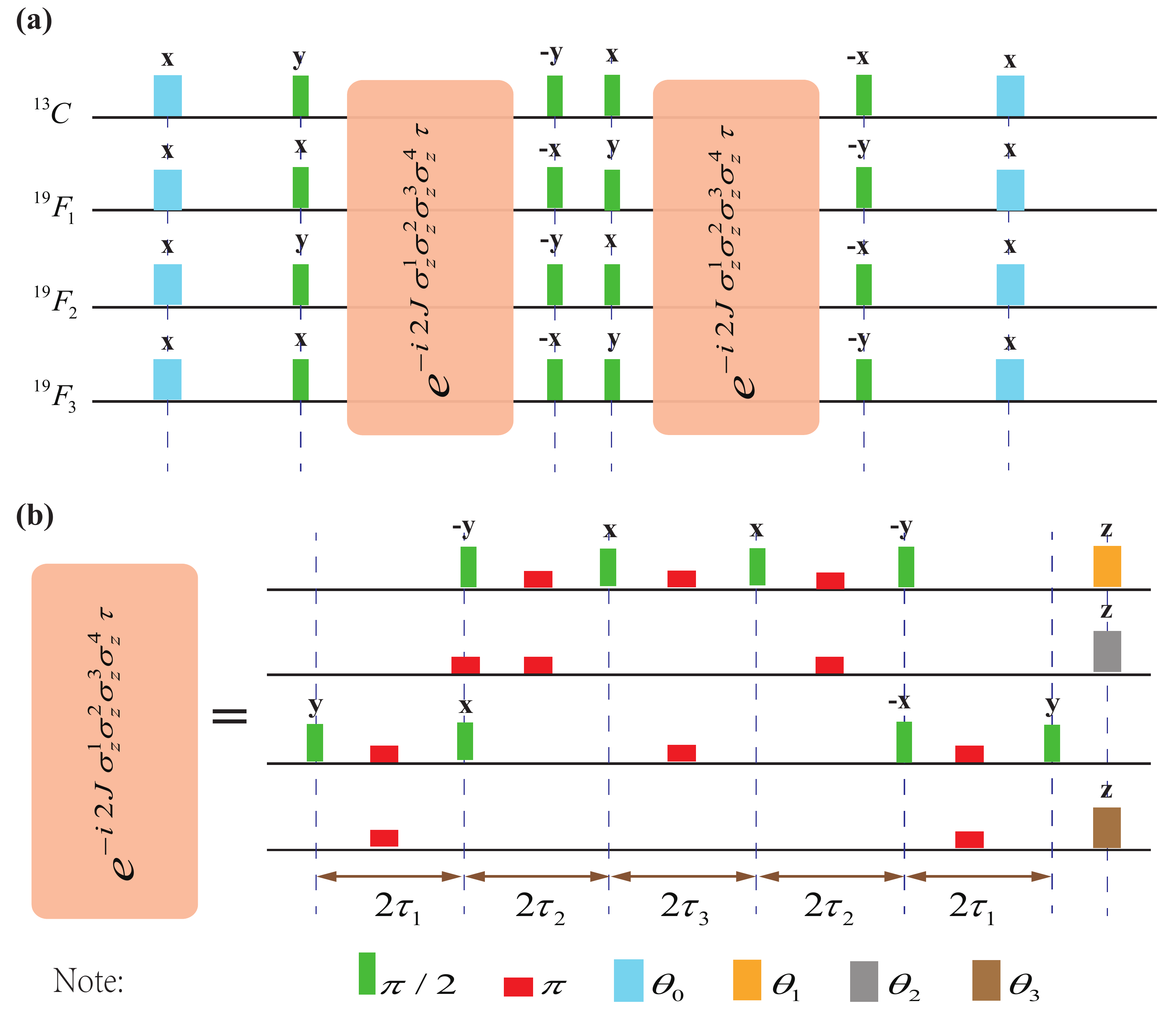}\newline
\caption{Pulse sequences for (a) simulating the transverse Wen-plaquette model of Eq. (1), and (b) four-body interaction, i.e., $\hat{\protect \sigma}_z^1\hat{\protect \sigma}_z^2\hat{\protect \sigma}_z^3\hat{\protect \sigma}_z^4$, where $\tau_1=1/4J_{34},\tau_2=1/4J_{12},\tau_3=2J\tau/\pi J_{13},\theta_0=-g\tau,\theta_1=-\omega_1/J_{34},\theta_2=-4\omega_2J\tau/\pi J_{13}$ and $\theta_3=\omega_4/J_{12}+4\omega_4J\tau/\pi J_{13}$.}
\label{pulseq}
\end{figure}

Using Trotter's formula, the target Hamiltonian (the transverse Wen-plaquette model in  Eq. (1)) can be created as an average Hamiltonian by concatenating evolutions with short periods
$$
e^{-i\hat{H}_{tol}\tau} = e^{-i\hat{H}_x\tau/2}e^{-i\hat{H}_{Wen}^4\tau}e^{-i\hat{H}_x\tau/2} + O(\tau^3),
$$
where $\hat{H}_x=-g\sum_{j=1}^4\hat{%
\sigma}_x^j$ and $\hat{H}_{Wen}^4=-2J(\hat{\sigma}_{1}^{x}\hat{\sigma}_{2}^{y}\hat{\sigma}_{3}^{x}\hat{\sigma}_{4}^{y}+\hat{\sigma}_{1}^{y}\hat{\sigma}_{2}^{x}\hat{\sigma}_{3}^{y}\hat{\sigma}_{4}^{x})$. This expansion faithfully represents the targeted evolution provided the duration $\tau$ is kept sufficiently short.  Due to $[ \hat{\sigma}_{1}^{x}\hat{\sigma}
_{2}^{y}\hat{\sigma}_{3}^{x}\hat{\sigma}_{4}^{y}, \hat{\sigma}_{1}^{y}\hat{\sigma}_{2}^{x}\hat{\sigma}_{3}^{y}\hat{\sigma}_{4}^{x}] =0 $,
\begin{eqnarray}
e^{-i\hat{H}_{Wen}^4\tau} = \bar{Y}_1 \bar{X}_2 \bar{Y}_3 \bar{X}_4  e^{ -i 2 J \hat{\sigma}_{1}^{z}\hat{\sigma}
_{2}^{z}\hat{\sigma}_{3}^{z}\hat{\sigma}_{4}^{z}  \tau}  Y_1 X_2 Y_3 X_4   \nonumber \\
  \cdot   \bar{X}_1 \bar{Y}_2 \bar{X}_3 \bar{Y}_4 e^{ -i 2 J \hat{\sigma}_{1}^{z}\hat{\sigma}
_{2}^{z}\hat{\sigma}_{3}^{z}\hat{\sigma}_{4}^{z}  \tau}   X_1 Y_2 X_3 Y_4. \nonumber
\end{eqnarray}
 Here the many-body interaction can be simulated by a combination of two-body interactions and RF pulses \cite{Tseng1999, Peng2009}:
\begin{align} \nonumber
&e^{-i 2J \sigma_z^1\sigma_z^2\sigma_z^3\sigma_z^4 \tau } \\ \nonumber 
&=e^{-i(\theta_1\hat{\sigma}_z^1+\theta_2\hat{\sigma}_z^2+\theta_3\hat{\sigma}_z^4)/2}Y_3e^{-i\hat{H}_{\text{NMR}}\tau_1}e^{-i\pi(\hat{\sigma}_y^3+\hat{\sigma}_y^4)/2} \\ \nonumber
&\cdot e^{-i\hat{H}_{\text{NMR}}\tau_1}\bar{Y}_1\bar{X}_3e^{-i\hat{H}_{\text{NMR}}\tau_2}e^{-i\pi(\hat{\sigma}_y^1+\hat{\sigma}_y^2)/2}e^{-i\hat{H}_{\text{NMR}}\tau_2} \\ \nonumber
&\cdot X_1e^{-i\hat{H}_{\text{NMR}}\tau_3}e^{-i\pi(\hat{\sigma}_x^1+\hat{\sigma}_x^3)/2}e^{-i\hat{H}_{\text{NMR}}\tau_3}X_1e^{-i\hat{H}_{\text{NMR}}\tau_2} \\ \nonumber
&\cdot e^{-i\pi(\hat{\sigma}_y^1+\hat{\sigma}_y^2)/2}e^{-i\hat{H}_{\text{NMR}}\tau_2}\bar{Y}_1X_3e^{-i\pi \hat{\sigma}_y^2/2}e^{-i\hat{H}_{\text{NMR}}\tau_1} \\ \nonumber
&\cdot e^{-i\pi(\hat{\sigma}_x^3+\hat{\sigma}_x^4)/2}e^{-i\hat{H}_{\text{NMR}}\tau_1}Y_3
\end{align}
Figure \ref{pulseq} shows the pulse sequences for simulating the transverse Wen-plaquette model of Eq. (1). The simulation method is in principle efficient
as long as the decoherence time is long enough.

\begin{figure}[tbp]
\centering \includegraphics[width=7cm]{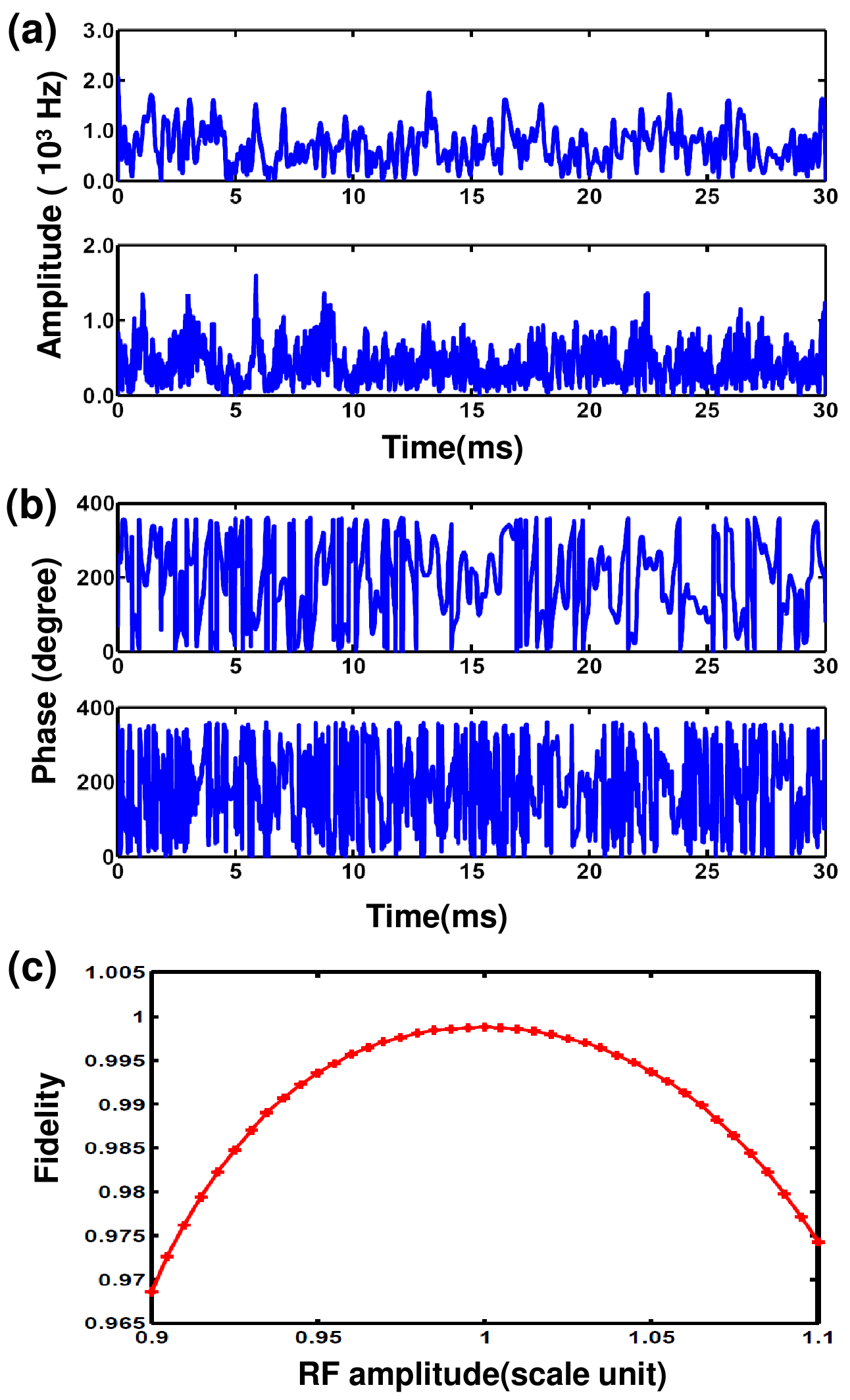}
\caption{An example of one GRAPE pulse for implementing adiabatic evolution.
Time-dependence of the amplitude (a) and the phase (b) of the GRAPE
pulse in the $^{13}$C (top) and $^{19}$F (bottom) channels. (c) Robustness
of the GRAPE pulse against RF inhomogeneities.}
\label{fig:grape}
\end{figure}

In order to overcome the accumulated pulse errors and the decoherence, we packed the adiabatic passage for each $J(m)$ ($m= 0, 1, 2, ...,M -1$) into one
shaped pulse calculated by the gradient ascent pulse engineering (GRAPE) method \cite{Glaser2005}, with the length of each pulse being 30 ms. All the pulses have theoretical delities over 0.995, and are designed to be robust against the inhomogeneity of radio-frequency pulses in the experiments. As an example, we show a GRAPE pulse in Fig.~\ref{fig:grape}.

\section{Experimental results and analysis}

\subsection{Experimental Spectra}

\begin{figure}[tbp]
\centering \includegraphics[width=9cm]{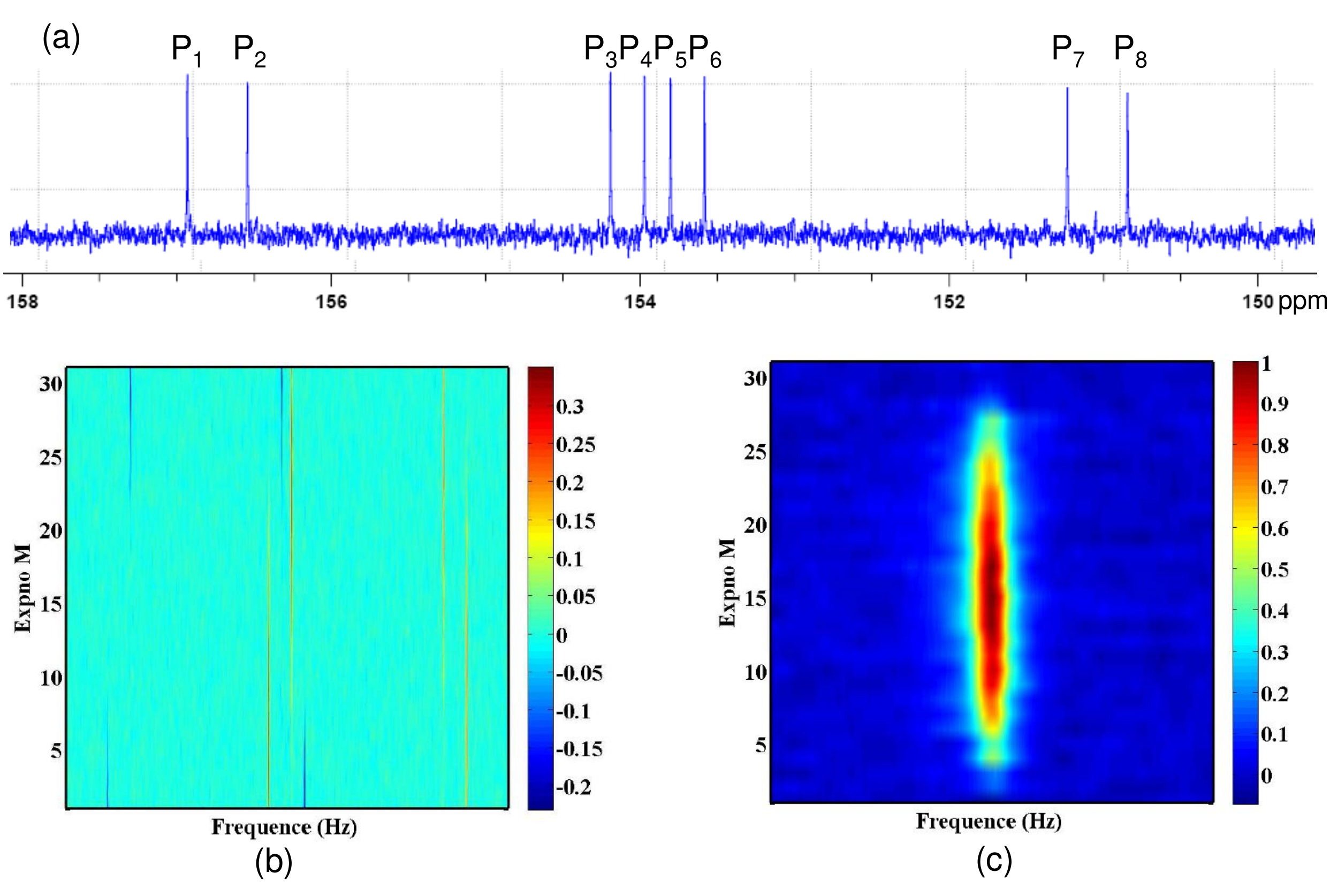}
\caption{Experimental $^{13}$C spectra for (a) equilibrium state after a reading-out pulse $[\frac{\pi }{2}]_{x}^{^{13}C}$,
(b) measuring the Wilson loop operator $\langle \hat{W}(C)\rangle$, and
(c) measuring the single-particle operator $P$. (b) and (c) are
two-dementional spectra with the total number of experiments $M=31$, and (c)
are the $^{19}$F-decoupled $^{13}$C spectra.}
\label{fig:spectrum2}
\end{figure}

Figure~\ref{fig:spectrum2} shows the experimental $^{13}$C spectra
for equilibrium state after a reading-out pulse $[\frac{\pi }{2}]_{x}^{^{13}C}$ (a), measuring the Wilson-loop operator $\langle \hat{W}(C)\rangle$ (i.e., after the reading-out pulse $[\frac{\pi }{2}]_{x}^{F_{1}}[\frac{\pi }{2}]_{\bar{y}}^{F_{2}}[\frac{\pi }{2}]_{x}^{F_{3}}$) and the single-particle operator $P$ (i.e., decoupling $^{19}$F without a reading-out pulse) on the $M = 31$ instantaneous states during the adiabatic passage, respectively.
The experimental values of $P$ were directly extracted from the integration of
the resonant peak of the $^{19}$F-decoupled $^{13}$C spectra, while the
experimental values of $\langle \hat{W}(C)\rangle$ determined by
\begin{equation}
\langle \hat{W}(C)\rangle=P_1-P_2-P_3-P_4+P_5+P_6+P_7-P_8
\end{equation}
where $P_i \, (i=1,2,\cdots,8)$ represents the integration of the $i^{th}$
resonant peak.

\subsection{State tomography of the ground states}

Due to the unlabeled sample, it is difficult to directly measure the $^{19}$F signals
related to quantum simulator with the $^{13}$C nucleus. Thus we transferred the states of the $^{19}$F spins to the $%
^{13} $C spin by a SWAP gate and read out the state information of the $^{19}$F spins through the $^{13}$C
spectra. To reconstruct the full density matrices of the four-qubit states,
we performed the 44 independent experiments (see Fig.~\ref%
{fig:tomography1}) to obtain the coefficients for all of the 256 operators which
comprise a complete operator basis of the four-qubit system. In the
experiment, this tomography involves 28 local operations and 3 SWAP gates.
All of these operations were realized by GRAPE pulses with 400 $\mu s$ for local
operations, 9 ms for the SWAP gates between $^{13}C$ and $F_1$, $F_2$ and 30
ms for the SWAP gate between $^{13}C $ and $F_3$ due to the relatively weak
coupling between them. Figure \ref{fig:DM_imag} shows some experimental
results for the ground states obtained during the adiabatic passage in the experiments.

\begin{figure}[tbp]
\centering \includegraphics[width=9cm]{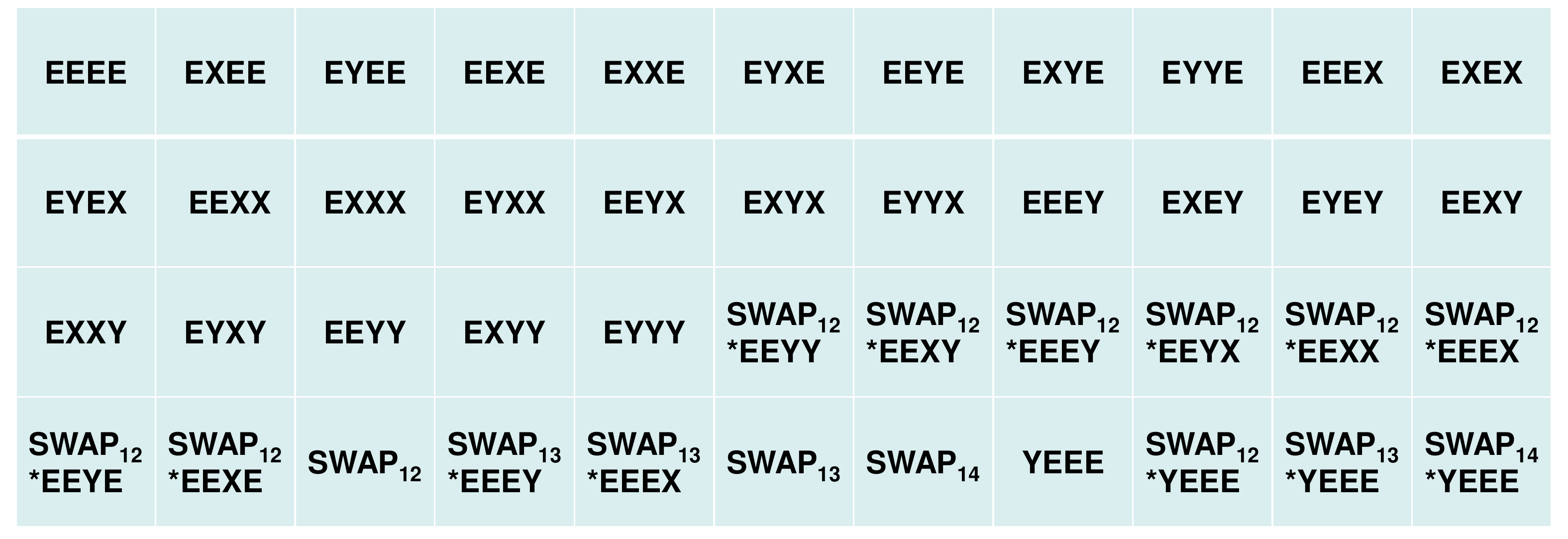}
\caption{Scheme of the reading-out pulses for the quantum state tomography for our four-qubit
quantum simulator. SWAP$_{ij}$ represents a SWAP gate between spin $i$ and $j
$ in order to transfer the $^{19}$F information to $^{13}$C, and then all
signals are obtained from the $^{13}$C spectra. }
\label{fig:tomography1}
\end{figure}

\begin{figure}[tbp]
\centering \includegraphics[width=9cm]{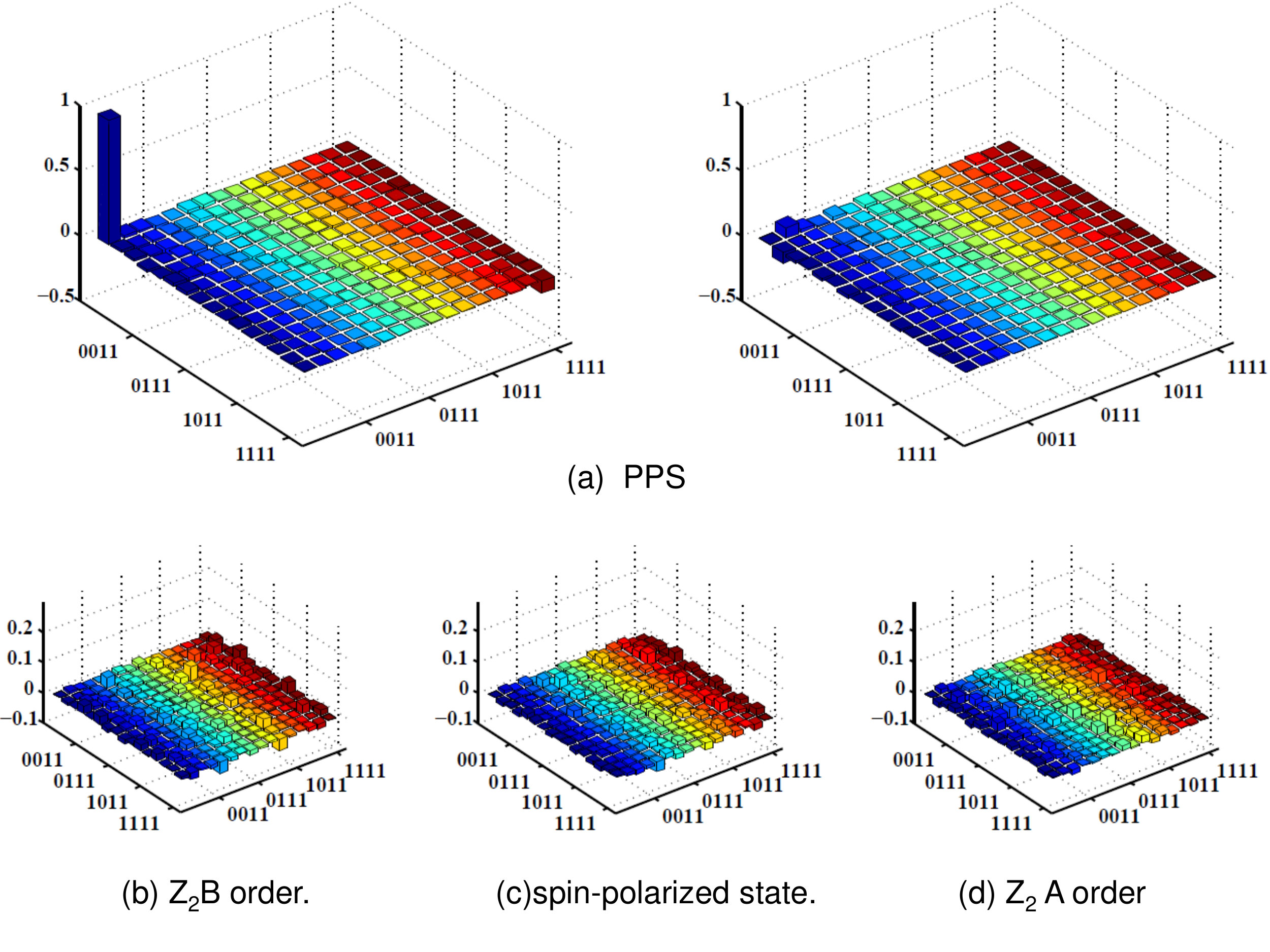}
\caption{Experimentally reconstructed density matrices. (a) Real (left) and imaginary (right) parts
of PPS, with the experimental fidelity around 97.7\%. (b)(c)(d)
represent the imaginary parts of $Z_2B$ order, spin-polarized state and $Z_2A
$ order, respectively (see Figure 4 in the main text for their real parts).}
\label{fig:DM_imag}
\end{figure}

\subsection{Error Analysis}

We calculated the standard deviations $\sigma = \sqrt{%
\sum_{i=1}^{M}(x_{exp}^i-x_{th}^i)^2 /M}$ for the experimental measurements
of the Wilson loop $\langle \hat{W}(C)\rangle$ and the single-particle
properties $P$. The results are listed in Table I. The standard deviations are small and
mainly caused by the imperfection of the initial-ground-state preparation,
the GRAPE pulses and the others which can be estimated by numerical
simulations. Taking the case with $g=1$ as an example, the simulated results
are also shown in Table I. Sim1 represents a numerical simulation where we
apply the theoretical GRAPE pulses $\hat{U}_{GRAPE}^{th}$ for the
adiabatic evolution on the idea initial state $\hat{\rho}_g^{th} = \vert
\psi_g^{th} \rangle \langle \psi_g^{th} \vert$, i.e., the simulated standard
deviations for $\langle \hat{W}(C)\rangle^{sim1} =Tr( \hat{\rho}_f^{sim1}
\hat{W}(C) )$ where $\hat{\rho}_f^{sim1} = \hat{U}_{GRAPE}^{th} \hat{\rho}%
_g^{th} \hat{U}_{GRAPE}^{th \dagger}$ and $P^{sim1} = \left| Tr[\hat{\rho}%
_f^{sim1} (\hat{\sigma}^1_x - i \hat{\sigma}^1_y) ] \right|$. This values
illustrate the errors only induced by the theoretical imperfection of GRAPE
pulses. Sim2 represents a numerical simulation where we apply $\hat{U}_{GRAPE}^{th}$
on the experimentally reconstructed density matrix of the initial state $\hat{\rho}_g^{exp} $, i.e., the simulated standard deviations for $\langle
\hat{W}(C)\rangle^{sim2} =Tr( \hat{\rho}_f^{sim2} \hat{W}(C) )$ where $\hat{%
\rho}_f^{sim2}= \hat{U}_{GRAPE}^{th} \hat{\rho}_g^{exp} \hat{U}_{GRAPE}^{th
\dagger}$ and $P^{sim2} = \left| Tr[\hat{\rho}_f^{sim2} (\hat{\sigma}^1_x -
i \hat{\sigma}^1_y) ] \right|$. This values account for the errors
contributed by the experimental imperfection of preparing initial ground
state. The remaining errors can come from the imperfections of experimental
quantum control, the static magnetic field and the spectral integrals and so
on.

\begin{table}[tbp]
\caption{The standard deviations of $\langle \hat{W}(C)\rangle $ and P for
the experiments and numerical simulations.}
\label{tab:error}
\begin{center}
\begin{tabular}{c|cc}
& $\sigma_{\langle \hat{W}(C)\rangle}$ & $\sigma_P$ \\ \hline
Exp & 0.091 & 0.078 \\
Sim1 & 0.043 & 0.038 \\
Sim2 & 0.066 & 0.043 \\
&  &
\end{tabular}
.
\end{center}
\end{table}


\begin{thebibliography}{99}

\bibitem{QPT} S. Sachdev, \emph{Quantum Phase Transition} (Cambridge University. Press, Cambrige 1999).

\bibitem{Landau1937}L. D. Landau, Phys. Zs. Sowjet \textbf{11}, 26 (1937).

\bibitem{Ginzburg1950} V. L. Ginzburg and L. D. Landau, J. Exp. Eheor.
Phys. \textbf{20}, 1064 (1950).

\bibitem{TSG8259} D. C. Tsui, H. L. Stormer, and A. C. Gossard, Phys. Rev. Lett. \textbf{48}, 1559 (1982);
R. B. Laughlin, \textit{ibid.}, \textbf{50}, 1395 (1983).

\bibitem{Wen1990TO}  X. G. Wen, Int. J. Mod. Phys. B \textbf{4}, 239 (1990).

\bibitem{Wen1990TO2} X.-G. Wen and Q. Niu, Phys. Rev. B \textbf{41}, 9377 (1990).

\bibitem{Arovas} D. Arovas, J. R. Schrieffer and F. Wilczek, Phys. Rev. Lett. \textbf{53}, 722 (1984).

\bibitem{Wen1995} X. G. Wen, Adv. Phys. \textbf{44}, 405 (1995).

\bibitem{k3} A. Kitaev and J. Preskill,  Phys. Rev. Lett. \textbf{96}, 110404 (2006); M. Levin and X. G. Wen, \textit{ibid.} \textbf{96}, 110405 (2006).

\bibitem{k1} A. Kitaev, Ann. Phys. (N.Y.) \textbf{303}, 2(2003).

\bibitem{Nayak2008} C. Nayak \textit{et al}., Rev. Mod. Phys. \textbf{80},1083 (2008).

\bibitem{Stern2013} A. Stern and N. H. Lindner, Science \textbf{339}, 1179-1181 (2013).

\bibitem{Wen2003} X. G. Wen, Phys. Rev. Lett. \textbf{90}, 016803 (2003).

\bibitem{k2} A. Kitaev, Ann. Phys. (N.Y.) \textbf{321}, 2(2006).

\bibitem{p3} J. Yu, S. P. Kou, and X. G. Wen, Europhys. Lett. \textbf{84}, 17004 (2008); S. P. Kou, J. Yu and X. G. Wen, Phys. Rev. B \textbf{80}, 125101 (2009).

\bibitem{HammaPRL2008} A. Hamma and D. A. Lidar, Phys. Rev. Lett. \textbf{100}, 030502 (2008).

\bibitem{HammaPRB2008} A. Hamma, W. Zhang, S. Haas, and D. A. Lidar, Phys. Rev. B \textbf{77}, 155111 (2008).

\bibitem{Duan2003} L. M. Duan, E. Demler, and M. D. Lukin, Phys. Rev. Lett. \textbf{91}, 090402 (2003); X. J. Liu, K. T. Law, and T. K. Ng, Phys. Rev. Lett. \textbf{112}, 086401 (2014).

\bibitem{Micheli2006} A. Micheli, G. K. Brennen, and P. A. Zoller, Nat. Phys. \textbf{2}, 341 (2006).

\bibitem{You2010} J. Q. You, X. F. Shi, X. D. Hu, and F. Nori, Phys. Rev. B \textbf{81}, 014505 (2010); L. B. Ioffe \textit{et al}., Nature \textbf{415}, 503-506 (2002).

\bibitem{Pan2012} X. C. Yao \textit{et al}., Nature \textbf{482}, 489-494 (2012); C. Y. Lu \textit{et al}., Phys. Rev. Lett. \textbf{102}, 030502 (2009); J. K. Pachos \textit{et al}., New. J. Phys. \textbf{11}, 083010 (2009).

\bibitem{Du2007} J. F. Du, J. Zhu, M. G. Hu, and J. L. Chen, arXiv:0712.2694v1 (2007); G. R. Feng, G. L. Long, and R. Laflamme, Phys. Rev. A \textbf{88}, 022305 (2013).


\bibitem{PRB82} X. Chen, Z. C. Gu, X. G. Wen, Phys. Rev. B \textbf{82}, 155138 (2010)

\bibitem{Feynman1982} R. P. Feynman, Int. J. Theor. Phys. \textbf{21}, 467 (1982).

\bibitem{Peng2005} X. H. Peng, J. F. Du, and D. Suter,  Phys. Rev. A \textbf{71}, 012307 (2005); K. Kim \textit{et al}., Nature \textbf{465}, 590 (2010); X. H. Peng, J. F. Zhang, J. F. Du and D. Suter,  Phys. Rev. Lett. \textbf{103}, 140501(2009); Gonzalo~A. \'Alvarez and Dieter Suter, Phys. Rev. Lett.  \textbf{104}, 230403 (2010).

\bibitem{Lu2011} J. F. Du \textit{et al}., Phys. Rev. Lett. \textbf{104}, 030502 (2010); B. P. Lanyon \textit{et al}. Nat. Chem. \textbf{2} 106 (2010); D. W. Lu \textit{et al}., Phys. Rev. Lett. \textbf{107}, 020501 (2011).

\bibitem{QSreview} I. M. Georgescu, S. Ashhab and F. Nori, Rev. Mod. Phys. \textbf{86}, 153 (2014).

\bibitem{Dennis2002} E. Dennis \textit{et al}., J. Math. Phys. \textbf{43}, 4452 (2002).

\bibitem{Jiang2008} L. Jiang \textit{et al}., Nat. Phys. \textbf{4}, 482-488 (2008).

\bibitem{WenPRD2003} X. G. Wen, Phys. Rev. D \textbf{68}, 065003 (2003).

\bibitem{SM} See Supplemental Material [url], which includes Refs. \cite{Feng2007,KouPRL2009,Claridge,Tseng1999}.

\bibitem{Feng2007} X. Y. Feng, G. M. Zhang, T. Xiang, Phys. Rev. Lett.
\textbf{98}, 087204 (2007).

\bibitem{KouPRL2009} S. P. Kou, Phys. Rev. Lett. \textbf{102}, 120402 (2009).

\bibitem{Claridge} Claridge, T. D. W. \emph{High resolution NMR techniques
in organic chemistry. Tetrahedron Organic Chemistry Series 19} (Elsevier,
Amsterdam, 1999).

\bibitem{Tseng1999} C. H. Tseng, et al, Phys. Rev. A \textbf{61}, 012302
(1999).

\bibitem{Peng2001} X. Peng \textit{et al}., Chem. Phys. Lett. \textbf{340} (2001).

\bibitem{Glaser2005} N. Khaneja \textit{et al}., J. Magn. Reson. \textbf{172}, 296 (2005).

\bibitem{Messiah1976} A. Messiah, Quantum Mechanics (Wiley, New York, 1976).

\bibitem{Kogut1979} J. B. Kogut, Rev. Mod. Phys. \textbf{51}, 659 (1979).

\bibitem{Lee} J. S. Lee, Phys. Lett. A \textbf{305}, 349--353(2002)

\bibitem{Wootters1998} W. K. Wootters, Phys. Rev. Lett. \textbf{80}, 2245 (1998).

\bibitem{Yu2009} J. Yu and S. P. Kou, Phys. Rev. B \textbf{80}, 075107 (2009).

\end{thebibliography}

\begin{thebibliography}{99}
\bibitem{Wen2003} X. G. Wen, Phys. Rev. Lett \textbf{90}, 016803 (2003).

\bibitem{Yu2008} J. Yu, S. P. Kou, \& X. G. Wen, Europhys. Lett., \textbf{84}%
, 17004 (2008).

\bibitem{Feng2007} X. Y. Feng, G. M. Zhang, T. Xiang, Phys. Rev. Lett.
\textbf{98}, 087204 (2007).

\bibitem{HammaPRL2008} A. Hamma, D. A. Lidar, Phys. Rev. Lett. \textbf{100},
030502(2008).

\bibitem{HammaPRB2008} A. Hamma, W. Zhang, S. Haas, D. A. Lidar, Phys. Rev.
B \textbf{77}, 155111 (2008).

\bibitem{k1} A. Kitaev, Ann. Phys. \textbf{303}, 2(2003).

\bibitem{KouPRL2009} S. P. Kou, Phys. Rev. Lett. \textbf{102}, 120402 (2009).

\bibitem{Yu2009} J. Yu and S. P. Kou, Phys. Rev. B \textbf{80}, 075107 (2009).

\bibitem{Claridge} Claridge, T. D. W. \emph{High resolution NMR techniques
in organic chemistry. Tetrahedron Organic Chemistry Series 19} (Elsevier,
Amsterdam, 1999).

\bibitem{Messiah1976} Messiah, A. \emph{Quantum Mechanics} (Wiley, New York,
1976).


\bibitem{Tseng1999} C. H. Tseng, et al, Phys. Rev. A \textbf{61}, 012302
(1999).

\bibitem{Peng2009} X. H. Peng, J. F. Zhang, J. F. Du and D. Sutter, Phys.
Rev. Lett \textbf{103}, 140501 (2009).

\bibitem{Glaser2005} Khaneja, N., Reiss, T., Kehlet, C., Schulte Herbr\"{u}%
ggen, T. \& Glaser, S. J., J. Magn. Reson \textbf{172}, 296 (2005).
\end{thebibliography}
\end{document}